\documentclass[a4paper,twoside]{article}

\usepackage{epsfig}
\usepackage{subcaption}
\usepackage{calc}
\usepackage{amssymb}
\usepackage{amstext}
\usepackage{amsmath}
\usepackage{amsthm}
\usepackage{multicol}
\usepackage{pslatex}
\usepackage{apalike}

\usepackage{url}
\usepackage{eurosym}

\usepackage{SCITEPRESS}     
\usepackage{tikz}
\usepackage[some,bottom]{background} 
\newcommand\copyrighttext{
 \footnotesize 
 Accepted at 2019 International Conference on Data Science, Technology and Applications (DATA) - \url{http://www.dataconference.org}.
 
 The final authenticated version will be available online in the conference proceedings - ISBN: 978-989-758-377-3, ISSN: 2184-285X.
 
 Copyright \textcopyright 2019 by SCITEPRESS – Science and Technology Publications, Lda. Personal use of this material is permitted.
  Permission from SCITEPRESS must be obtained for all other uses, in any current or future 
  media, including reprinting/republishing this material for advertising or promotional 
  purposes, creating new collective works, for resale or redistribution to servers or 
  lists, or reuse of any copyrighted component of this work in other works. 
 }
\newcommand\copyrightnotice{%
\begin{tikzpicture}[remember picture, overlay]
\node[yshift=10pt]
{\fbox{\parbox{\dimexpr\textwidth-\fboxsep-\fboxrule\relax}{\copyrighttext}}};
\end{tikzpicture}
}
\backgroundsetup{opacity=1, scale=1, angle=0, contents={
\copyrightnotice
}}
\SetBgOpacity{1}
\SetBgScale{1}
\SetBgAngle{0}
\SetBgContents{\copyrightnotice}
\SetBgColor{black}

\begin{document}

\title{Application of Data Stream Processing Technologies in Industry 4.0 - What is Missing?}

\author{
\authorname{Guenter Hesse\sup{1}\orcidAuthor{0000−0002−7634−3021}, 
Werner Sinzig\sup{1}, 
Christoph Matthies\orcidAuthor{0000−0002−6612−5055} 
and 
Matthias Uflacker\sup{1}
}
\affiliation{\sup{1}Hasso Plattner Institute, University of Potsdam, August-Bebel-Str. 88, 14482 Potsdam, Germany}
\email{firstname.lastname@hpi.de}
}


\abstract{
Industry 4.0 is becoming more and more important for manufacturers as the developments in the area of Internet of Things advance. 
Another technology gaining more attention is data stream processing systems.
Although such streaming frameworks seem to be a natural fit for Industry 4.0 scenarios, their application in this context is still low.
The contributions in this paper are threefold.
Firstly, we present industry findings that we derived from site inspections with a focus on Industry 4.0. 
Moreover, our view on Industry 4.0 and important related aspects is elaborated. 
As a third contribution, we illustrate our opinion on why data stream processing technologies could act as an enabler for Industry 4.0 and point out possible obstacles on this way. 
}

\onecolumn \maketitle \normalsize \setcounter{footnote}{0} \vfill
\BgThispage

\section{\uppercase{Introduction}}
\label{sec:introduction}

\noindent In the backlight of technological and economic developments, the term Industry 4.0 gained more and more popularity. 
Technically, new Internet of Things (IoT) technologies, such as sensors, are being created, sensor accuracy increases and analytical IT systems are being developed that allow querying huge amounts of data within seconds, to name but a few.
On the economic side, a substantial price decrease for sensor IoT equipment can be recognized. 
This trend is expected to continue in the following years.
To be more concrete, the price for an IoT node is expected to drop by about 50\% from 2015 to 2020~\cite{wee2015industry}. 
These developments fostered the increased deployment of IoT technologies in companies, especially in the manufacturing sector, and thus, more IoT data  is available to companies~\cite{economist}.
Monetarily expressed, the total global worth of IoT technology is expected to reach USD 6.2 trillion by 2025.
One of the industry sectors investing most on IoT is industrial manufacturing~\cite{intel}.

A related term in the context of manufacturing that gained attention in the past years is Industry 4.0. 
One reason for that is the potential that is seen in it with respect to creating an added value for enterprises.
A survey conducted by McKinsey in January 2016 amongst enterprises in the US, Germany, and Japan with at least 50 employees highlights the significance of Industry 4.0. 
The study reveals, e.g., that the majority of companies expect Industry 4.0 to increase competitiveness~\cite{bauer2016industry}.


One of the identified key challenges is integrating data from different sources to enable Industry 4.0 applications~\cite{bauer2016industry}.
Especially with the emerging significance of IoT data, the fairly old challenge of integrating disparate data sources gets a new flavor. 
Data Stream Processing Systems (DSPSs) can be a technology suitable for tackling this issue of data integration.
Within this paper, a view on Industry 4.0 as well as the potential of Data Stream Processing technologies in that context is presented.

Following the introduction, industry insights related to Industry 4.0 observed through interviews and site inspections are highlighted.
In Section~\ref{sec:industry4.0}, we elaborate our view on Industry 4.0, i.e., our definition as well as our view on data integration and IoT.
Afterward, Section~\ref{sec:dsps} discusses DSPSs and their role in the area of Industry 4.0, including challenges regarding their application in Industry 4.0 settings. 
A section to related work and a conclusion complete this paper.


\section{\uppercase{Observations in Industry}}
\label{sec:obervations}

\noindent Beginning in 2015, we conducted interviews with multiple enterprises with a focus on Industry 4.0 implementation strategies and associated challenges and solutions.  
In this section, we describe and contrast the Industry 4.0 efforts of two selected companies.
Both enterprises belong to the manufacturing sector and are comparatively large, with more than 10,000 employees and revenue of more than \EUR{1}bn each.

\subsection{Company I}
The first company collects sensor and log data from two sources, its machines used for manufacturing as well as from its sold products used by its customers.
About 250 of the vended machines were configured to collect and send sensor and log data to an external central cloud storage service back in late 2015.
The data is sent as a batch every 23 hours and includes several state values,  such as temperature and position information.
Overall, that results in about 800GB data on a monthly basis.
Another external company is responsible for data cleansing and some basic calculations.
The results are then used by Company I.
As Company I is producing the machines, they also developed the format of the log data that is collected. 
Over time, this format changed with different software releases, which introduces additional complexity with respect to data integration. 

Regarding the machines used for manufacturing, five machines are configured to collect sensor data.
This data is recorded every 100ms and sent every hour to the same cloud storage service.
Each batch is about 20MB big with respect to size.
It contains, e.g., information about energy consumption and position data. 

As of late 2015, none of the collected data has ever been deleted.
Moreover, the stored sensor data had not found its way into an application at that point in time. 
However, Company I expected growth in its services area.
As part of that, it could imagine offering several services around its products for which the collected data would be useful.
Predictive maintenance or remote service and support scenarios are an example of such services. 
Besides, the collected data could reveal further insights about product usage and behavior, which could help product development. 
The internally captured data could be used for, e.g., predictive maintenance or quality improvements scenarios. 
The knowledge about production behaviors of previously manufactured products can be combined and gained learnings can be used to support product development and production planning.

\subsection{Company II}

\noindent The second company has several measurement stations in its production line.
At these stations, certain parts of the product-to-be are gauged.
Resulting data are mostly coordinates, e.g., borehole positions.
By doing so, possible inaccuracies added in previous production steps are identified.
If an inaccuracy exceeds a threshold, the corresponding product is removed from the production line and the mistakes are corrected if possible.

Furthermore, there is a central database storing all warning or error messages that appear in the production line. 
A higher five-digit number of messages occurs on a single day on average, whereas this number can go up to more than a million messages.
Besides the time stating when the deviation took place, the point in time when it is remedied is stored next to further values describing the event.

With respect to Industry 4.0 applications, the company was in the evaluation process, meaning thinking about how the existing data could be used for such scenarios. 
Back in 2015, the stored warnings and errors had a documentary character rather than being used in applications for, e.g., preventing future deviations or optimizing processes. 
However, it was an objective to leverage this data more in such kind of programs. 
The measurement data was considered first for this kind of evaluations.

\subsection{Industry Study Conclusions}
Both presented and studied companies have in common the positive view on Industry 4.0, meaning they see it as a chance rather than a threat, which fits the before-mentioned survey conducted by McKinsey~\cite{wee2015industry}.
However, neither of the companies, which can both be considered as leaders with respect to market share or revenue, have been able to significantly leverage the potential of Industry 4.0.
None of them is using data stream processing technologies in this domain so far.
To be more concrete, IoT data is collected but no major new applications using this data or even combining it with business data have been introduced. 
That might serve as an example for technological leaders struggling to implement new innovations.
This situation is often referred to as the innovator's dilemma, which elaborates on the challenge for successful companies to stay innovative~\cite{christensen2013innovator}. 

\section{\uppercase{Industry 4.0}}
\label{sec:industry4.0}

\noindent In this section, we elaborate on the Industry 4.0-related topics of data integration and the Internet of Things.
Based on that, we present our view on the term Industry 4.0 afterward.

\subsection{Data Integration}

Being able to map data from different sources belonging together is crucial to get holistic pictures of processes, entities, and relationships.
The more data can be combined, the more complete and valuable is the created view that is needed for fact-based assessments and decisions within enterprises. 
Consequently, better data integration and so more available data can lead to greater insights and understanding, better decisions, and thus, to a competitive advantage.

After giving a brief overview of the current situation in enterprises that we discovered through conducted interviews, site inspections, and research projects, our views on the terms horizontal and vertical data integration are elaborated.

\subsubsection{Current Situation in Enterprises}

Business processes are central artifacts that describe an enterprise and the infrastructure they are embedded in. 
Business systems represent such processes digitally, e.g., in the form of data model entities like a customer, customer order, product, production order, or journal entries. 
For different companies, the semantics of these entities can vary, which can hamper data integration exceeding company boundaries.
Within a single company, definitions should be clear.
However, that does not necessarily represent reality.

Besides business systems, sensors or IoT-related technologies become a greater source of data that is describing processes and infrastructure in an enterprise. 
This information is usually connected to business systems, such as an Enterprise Resource Planning (ERP) system or alike, via a Machine Execution System (MES) in the manufacturing sector.
Additionally, there might be more systems installed underlying the MES responsible for managing the shop floor.

A typical IT landscape comprises many different business systems. 
D\"{o}hler, for instance, a company with more than 6,000 employees from the food and beverage industry, has more than ten business systems and supporting systems that need to be managed and where ideally data can be exchanged amongst each other.
In addition to an ERP system, there are, e.g., systems for customer relationship management, extended warehouse management, and an enterprise portal~\cite{doehler1,doehler2}. 
Often, fragmented IT landscapes have been developed historically and complexity increased through, e.g., acquisitions. 
Simplification is a challenge in companies, e.g., due to the lack of knowledge about old systems that might be still used. 

But even if all business systems are from the same vendor, entities can differ between systems.
A centralization to a single ERP system is unlikely to happen for multiple reasons.
Such arguments can be related to aspects like data security of sensitive data, e.g., HR data shall be decoupled from the main ERP, or the wish to be not dependent on a single software vendor for economic or risk diversification reasons.

\begin{figure*}[ht]
\centering
\includegraphics[width=0.9\textwidth]{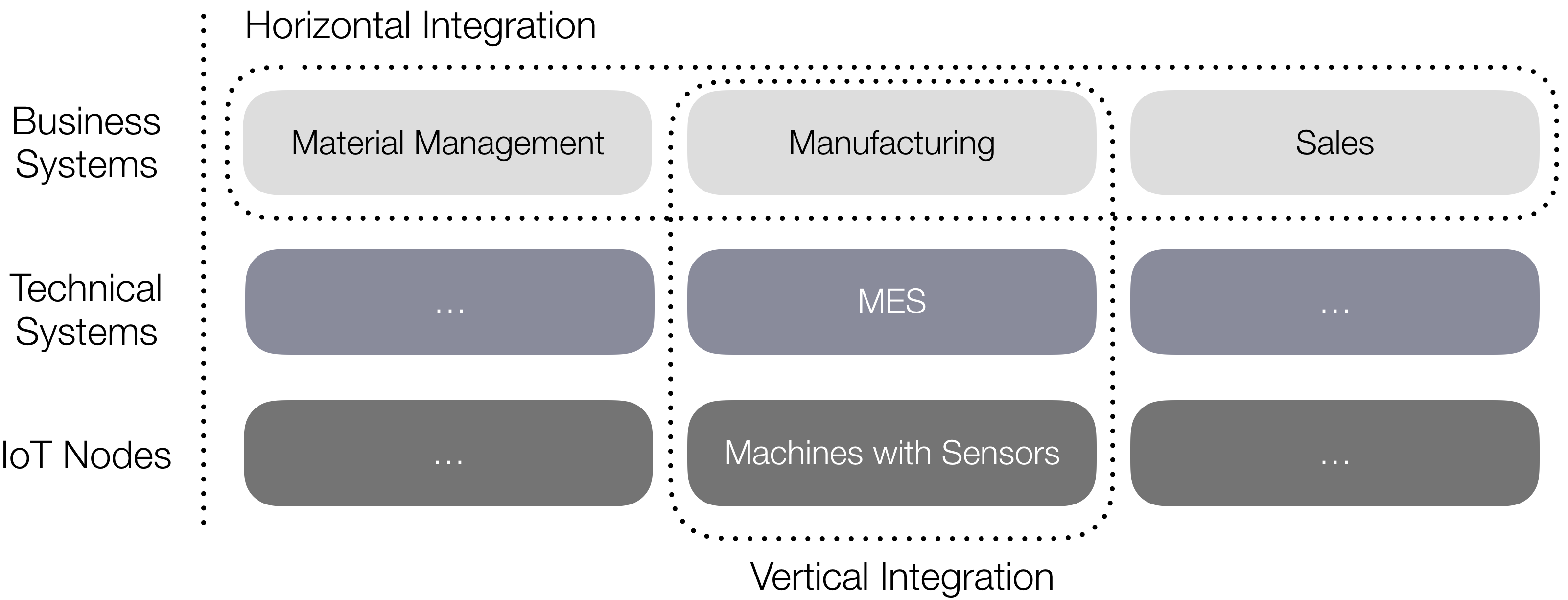}
\caption{Conceptual Overview of Data Integration in the Context of Industry 4.0}
\label{fig:dataintegration}
\end{figure*}

Figure~\ref{fig:dataintegration} visualizes a very simplified IT landscape how it can be found at companies belonging to the manufacturing sector. 
It distinguishes between different system categories and highlights the areas of horizontal and vertical integration, that are explained in Section~\ref{sec:horizontal} and Section~\ref{sec:vertical} respectively.

\subsubsection{Horizontal Data Integration}
\label{sec:horizontal}
We see horizontal data integration as a holistic view of business processes, i.e., from the beginning to the end. 
Technically, that means joining database tables stored within business systems that are involved in the business process execution as conceptually outlined in Figure~\ref{fig:dataintegration}.
These links can be established, e.g., through foreign key dependencies. 
The greater the number of tables that can be connected, the more detailed and valuable the resulting view on a process.
As mentioned before, enterprises generally have multiple business systems for the elaborated reasons, which increases the effort for achieving a horizontal data integration.
Compared to vertical integration, horizontal integration is further developed having relatively advanced software solutions for achieving it. 

\subsubsection{Vertical Data Integration}
\label{sec:vertical}
Vertical data integration describes the connection of technical data created by IoT technologies and business systems, including the systems in between these two layers as depicted in Figure~\ref{fig:dataintegration}.
That means, two different kinds of data have to be combined in contrast to integrating only business data as in horizontal data integration. 
These distinct data characteristics introduce new challenges.

While business data is well-structured and with a comparatively high degree of correctness, sensor data can be relatively unstructured and error-prone.
Contrary to the close business process reference of business data, sensors have a strong time and location reference.
Moreover, both volume and creation speed of IoT data is generally higher, which impacts, e.g., the performance requirements on IT systems handling this kind of data~\cite{DBLP:conf/tpctc/HesseRMLKU17}.

Moreover, it is a challenge to map entities in business processes, such as a product that is being produced, to the corresponding IoT data that has been measured while exactly this product has been produced at the corresponding workplace. 
In contrast to integrating relatively homogeneous data among business systems, foreign keys cannot simply be used.
Instead, a time-based approach is often applied, which can potentially introduce errors due to imprecise time measurements.
However, the progress of vertical data integration we experienced in site inspections, e.g., in the form of being able to map sensor measurements created at production machines to the corresponding products that were being produced, is not as advanced as the horizontal integration.
Nevertheless, vertical integration as in the previously described scenario is desired since it can help to get further insights about processes and thus, support to create an added value. 



\subsection{Internet of Things}
\label{sec:iot}
Internet of Things is a term often used in the context of Industry 4.0 that has versatile meanings. 
Originally emerged out of the area of radio frequency identification (RFID), where it described connected physical objects using this technology, the term IoT became broader over recent years.  
It is not limited to RFID technology anymore, but also comprises, e.g., things connected via sensors or machine-to-machine communication.
Additionally, applications leveraging these technologies are referred to as IoT~\cite{ashton2009internet,DBLP:journals/adhoc/MiorandiSPC12}. 

We see IoT as network-connected physical objects, whereas it does not matter which exact technology is used for establishing a connection.
Moreover, IoT is an enabler for Industry 4.0 as it is driving vertical data integration and thus, paving the way for new business applications.
Through making machines or physical objects in general digitally accessible, new data can be analyzed, new insights be gained and a more holistic view of processes can be created.
This increased level of live information can lead to a competitive advantage for enterprises. 

\subsection{Our View on Industry 4.0}

We see Industry 4.0 as a term describing an advanced way of manufacturing enabled and driven by technological progress in various areas. 

These areas can be categorized into two groups, developments with respect to IoT technologies and regarding IT systems.
While the advances related to IoT enable to gain new, higher volumes and more precise measurements, the IT system development progresses allow to analyze high volumes of data with reasonable response times nowadays. 
Moreover, high volumes of data created with a high velocity can also be handled with the help of modern DSPSs.

These achievements lead to new opportunities in manufacturing.
New data is being generated in high volume and velocity, which can now also be analyzed in a reasonable amount of time with state-of-the-art IT systems.
This natural fit of two technological developments generates opportunities.
Making use of both advances in combination with full data integration, i.e., horizontally as well as vertically, raises the level of detail and completeness enterprises can have on their processes and entities.
This information gain  
\begin{itemize}
    \item leads to the enablement of better data-driven decisions,
    \item facilitates new insights into processes or entities,
    \item creates the opportunity for new business applications, and
    \item allows for rethinking the way of manufacturing.
\end{itemize}
 
Specifically, holistic data integration enables a flexible and more customizable production, i.e., moving from a nowadays commonly existing batch-wise production to piece-wise production while not sacrificing economic performance.
Although we have not observed a batch size of one as an explicitly formulated objective in our side inspections, it was considered as a desirable situation.
Generally, we got the impressions that there are greater challenges related to IT compared to those related to the engineering aspect of IoT. 

\section{\uppercase{Data Stream Processing Systems}}
\label{sec:dsps}

\noindent In this section, our view on the potential role of DSPSs in the context of Industry 4.0 is presented.
Moreover, the related challenges that need to be tackled are highlighted.

\subsection{A Possible Role in Industry 4.0}

Although data stream processing systems is not a new technology, it gained more attraction in the past couple of years~\cite{DBLP:conf/icpads/HesseL15}. 
Reasons for that are technological advances, e.g., with respect to distributed systems on the one hand, and on the other hand the grown need for such systems due to the increased data masses that are being created through developments like IoT for instance.

We think that stream processing technologies have the potential to play a central role in the context of Industry 4.0.
A reason for that is its suitability regarding the data characteristics of processed data, which fit the overall purpose behind DSPSs.
Instead of issuing a query that is executed once and returns a single result set as in a database management system (DBMS), DSPSs execute queries on a continuous basis.
Similarly, IoT data is often generated on a continuous basis, which is contrary to traditional business data.

Altering requirements, e.g., due to growing data volumes introduced by added machines or advanced IoT technologies, can be handled as modern DSPSs are typically scalable distributed systems.
As another consequence, high elasticity is enabled, i.e., nodes can be added or removed from the cluster as the workload increases or decreases. 
This flexibility is advantageous from an economic perspective. 
Especially manufacturers that do not produce during certain periods, generally speaking companies with large IoT workload variations, can benefit. 

\begin{figure}[ht]
\centering
\includegraphics[width=0.358\textwidth]{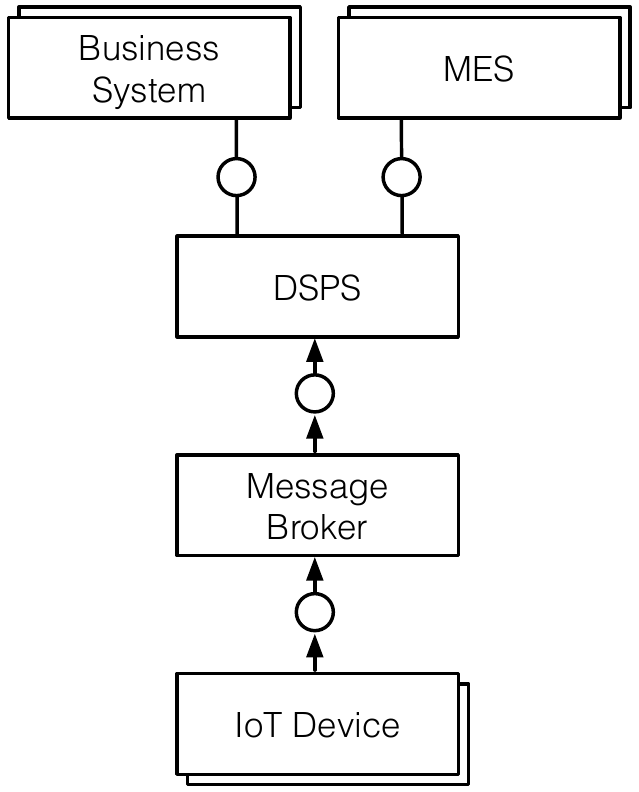}
\caption{Example Industry 4.0 IT Architecture in Fundamental Modeling Concepts (FMC)}
\label{fig:arch}
\end{figure}

Scalability can be reached by using a message broker between the sources of streaming data and the DSPS.
That is a common approach seen in many architectures, both in industry and science~\cite{DBLP:conf/tpctc/HesseRMLKU17}.
A schematic overview of a possible architecture is visualized in Figure~\ref{fig:arch}.

IoT devices, such as manufacturing equipment, can send their measurements to a message broker, from which a DSPS can consume the data.
Streaming applications that require more than IoT data, i.e., programs that need vertical integration, can also be realized using DSPSs.
Corresponding data can be consumed via established interfaces, such as JDBC, and enrich the IoT data.
If a horizontal data integration can be achieved in the business system layer.
A holistic view on entities or processes can then be created in the DSPS where all data is brought together.
Additionally, data from MES systems or alike can be integrated as depicted in Figure~\ref{fig:arch}.
That makes data stream processing technologies a suitable framework for developing Industry 4.0 applications whose use case does not have further requirements that can not be satisfied in this setting.

Summarizing, since DSPSs are capable of handling the high volume and high-velocity IoT data as mentioned previously, they can act as an enabler for vertical integration and thus, for Industry 4.0 scenarios.
Data can be analyzed on the fly without the need of storing high volumes of data in advance, which has a positive impact economically as well as on the performance side.
An imaginable pre-aggregation that would lower these effects is not needed.
Moreover, aggregation comes at the cost of data loss and thus, sacrifices accuracy. 

\subsection{Challenges}

Certain challenges exist that could hinder an establishment of data stream processing technologies in the context of Industry 4.0 on a broader scale.

One reason is the existing lack of a broadly accepted abstraction layer for formulating queries or developing applications, such as SQL for DBMSs.
Similarly, Stonebreaker, \c{C}etintemel and Zdonik mentioned the need for DSPSs to support a high-level stream processing SQL as one of the eight requirements of real-time stream processing they defined in~\cite{DBLP:journals/sigmod/StonebrakerCZ05}.
The lack of such an established abstraction layer introduces multiple challenges.
It reduces flexibility for enterprises as after choosing a certain system, the boundaries to exactly this system are comparatively tight. 
Switching to another framework, e.g., due to altered system requirements or changed performance ratios amongst the group of existing systems, is more complex and thus, costlier for companies.
Streaming applications need to be developed using native system APIs, which results in high porting effort if a system is supposed to be exchanged compared to the effort needed for switching a DBMS.
The resulting potentially needed SQL adaptions are relatively small since the same abstraction layer, namely SQL, is also used in the new system in typical scenarios.
There are multiple system-specific SQL dialects developed for stream processing frameworks, but none of them gained broader acceptance.
However, there is the open-source project Apache Beam aiming to close this gap.
It is not a domain-specific language like SQL, but a software development kit that allows writing programs, which can be executed on any of the supported stream processing engines.
The impact of using this abstraction layer on selected state-of-the-art DSPSs with respect to performance is analyzed in~\cite{icdcshesse}.

Furthermore, identifying the most suitable system might be a challenge for enterprises. 
Although the growing number of DSPSs that have been developed in recent years is generally a good thing, the more choice the harder to make a decision for choosing a system.
Typically, performance benchmarks are used for this task. 
Similarly to the previously described circumstances regarding the abstraction layers, the situation for DBMSs is more sophisticated.
While there are many well-known and often used benchmarks for databases, such as TPC-C, TPC-H, or TPC-DS, the area of DSPS benchmarks is significantly less developed.
Linear Road is probably the best-known benchmark for stream processing architectures~\cite{DBLP:conf/vldb/ArasuCGMMRST04}.
However, it does not reflect typical Industry 4.0 scenarios in contrast to a benchmark currently under development and proposed by~\cite{DBLP:conf/debs/HesseMRU17}, which could close the gap of not having a suitable benchmark for comparing different DSPSs for use in the Industry 4.0 domain.

Another challenge we recognized in site inspections is the identification of Industry 4.0 scenarios that possibly create an added value.
Although this situation is not directly linked to DSPSs, thoughts about Industry 4.0 and technologies that can be used barely include stream processing frameworks and their capabilities based on our industry experiences. 
This lack of awareness of streaming technologies results in not considering it for new application scenarios.
Moreover, when taken into account, there are often reservations, such as that there is no or only little knowledge about these technology amongst the employees.
Another fear is that modern DSPSs are very complex systems, which are hard to maintain and difficult to use for application development.
However, these points could be eliminated automatically in near future if development efforts and improvements of DSPSs stay as high as they are at the moment. 

\section{\uppercase{Related Work}}

\noindent A recent work developed a framework called Production Assessment 4.0, which aims to support enterprises developing Industry 4.0 use cases.
For doing so, they made use of the design thinking approach.
After elaborating on the framework and its processes, a section about its evaluation is presented.
Production Assessment 4.0 was evaluated in several consulting projects with enterprises.
However, no details about, e.g., their data characteristics or their state of Industry 4.0 adoption progress are given~\cite{bauer2018production}.

With respect to Industry 4.0, there are many existing definitions and views published.
An overview of selected perceptions of Industry 4.0 is presented in~\cite{MRUGALSKA2017466}.
Moreover, it also states that there is no generally accepted definition for the term Industry 4.0.

The Association of German Engineers (VDI) published an architecture reference model for Industry 4.0 named RAMI 4.0~\cite{hankel2015reference}.
It comprises three different dimensions, namely hierarchy levels in factories, the product life cycle value stream, and an architecture dimension containing several layers from physical things up to the organization and business processes.

Another work presents design principles for Industry 4.0 that are derived through text analysis and literature studies~\cite{DBLP:conf/hicss/HermannPO16}. 
Thereby, it is aimed to help both, the scientific community and practitioners with this result.
In total, four design principles were identified, namely technical assistance, interconnection, decentralized decisions, and information transparency. 

With regard to challenges, there is the previously mentioned work by Stonebreaker, \c{C}etintemel and Zdonik that defines eight requirements for real-time stream processing.
Particularly, these are the need (1) to keep the data moving, (2) for a streaming SQL language as highlighted before, (3) for the ability to handle stream imperfections, (4) to generate predictable outcomes, and (5) to integrate stored as well as streaming data, which fits to the Industry 4.0 scenarios where business (stored) and IoT (streaming) data are integrated. 
Furthermore, the requirement (6) to ensure data safety and availability, (7) to automatically scale and partition programs, and (8) to process and respond immediately are highlighted.

\section{\uppercase{Conclusion}}
\label{sec:conclusion}

\noindent The present paper pictures a point of view on Industry 4.0 and on data stream processing systems in its context.
Thereby, contributions are threefold.
First, we present insights about current situations and opinions at two selected companies with respect to Industry 4.0.
This includes information about data characteristics and Industry 4.0 applications.
All findings were derived from site inspections and alike.

Secondly, a viewpoint on Industry 4.0 as well as on further important and closely related aspects is given.
Among others, it ensures a common understanding needed for the third contribution.

This third part is about data stream processing systems. 
Particularly, it is about why and how this technology could become an enabler for Industry 4.0.
A possible architecture for Industry 4.0 scenarios is proposed and obstacles hindering DSPSs from being applied more in this context are pointed out.

\bibliographystyle{apalike}
{\small
\bibliography{example}}

\end{document}